\documentclass[prb,twoside,twocolumn,showpacs]{revtex4}
\usepackage{graphicx}
\usepackage{amsmath}
\usepackage{amssymb}

\newcommand{\cpd}{Pr$_6$Ni$_2$Si$_3$}
\newcommand{\cpdf}{Pr$_5$Ni$_2$Si$_3$}
\newcommand{\cpdft}{Pr$_{15}$Ni$_7$Si$_{10}$}

\newcommand{\ca}{\emph{c}-axis}
\newcommand{\ML}{$M_{\rm L}$}
\newcommand{\MT}{$M_{\rm T}$}
\newcommand{\MX}{$M_{\rm X}$}
\newcommand{\MY}{$M_{\rm Y}$}

\begin{document}

\title{Exotic (anti)ferromagnetism in single crystals of \cpd}
\author{Y. Janssen}
\email[Present address:] {Brookhaven National Laboratory, Upton, NY
11973, USA; yjanssen@bnl.gov}
\author{K. W. Dennis}
\author{R. Prozorov}
\author{P.~C. Canfield}
\author{R. W. McCallum}
\affiliation{Ames Laboratory DOE and Department of Physics and
Astronomy, Iowa State University,Ames,IA 50011,USA}

\date{\today}
\begin{abstract}
The ternary intermetallic compound \cpd, is a member of a structure series of compounds based on 
a triangular structure where the number of Pr atoms in the prism cross section can be 
systematically varied.  \cpd\ contains two distinct Pr lattice sites which result in complex 
interactions between the magnetic ions.  Extensive measurements of specific heat and 
magnetization on single crystal samples indicate that \cpd\ orders with both a ferromagnet 
and an antiferromagnet component, with ordering temperatures of  39.6 K and $\sim$ 32 K, 
respectively. The ferromagnetic component // \ca\ is accompanied by a large hysteresis, 
and the antiferromagnetic component, $\bot$ \ca\ is accompanied by a spin-flop-type 
transition. More detailed measurements, of the vector magnetization, indicate that the 
ferromagnetic and the antiferromagnetic order appear independent of each other. These 
results not only clarify the behavior of \cpd\ itself, but also of the other members 
of the structure series, \cpdf\ and \cpdft.
\end{abstract}


\maketitle

\section{Introduction}
\label{intro}

A magnetic system with magnetic moments on non-equivalent crystallographic sites may be 
difficult to analyze experimentally.  
Different site symmetries and interatomic spacings can produce magnetic interactions with 
different signs or strengths as well as different crystalline electric field splittings and 
even valencies. 
If, however, the considered magnetic system forms part of a series, similarities in magnetic 
properties of the members may lead to a greater understanding of the properties of all members. 
Rare-earth intermetallic compounds often form natural series, because, due to the chemical
similarity of rare earths, a particular rare earth can be replaced for another, and the 
resulting magnetic systematics can be appreciated (See e.g.\ Ref.\onlinecite{Taylor71}).
Another type of series that may be considered is the structure series. In a structure series, 
structural features are systematically repeated, which may help in understanding the physical 
properties of members of such series. 

\begin{figure}[!ht] 
\includegraphics[width=0.45\textwidth,trim=0in 0in 0.0in 0in]{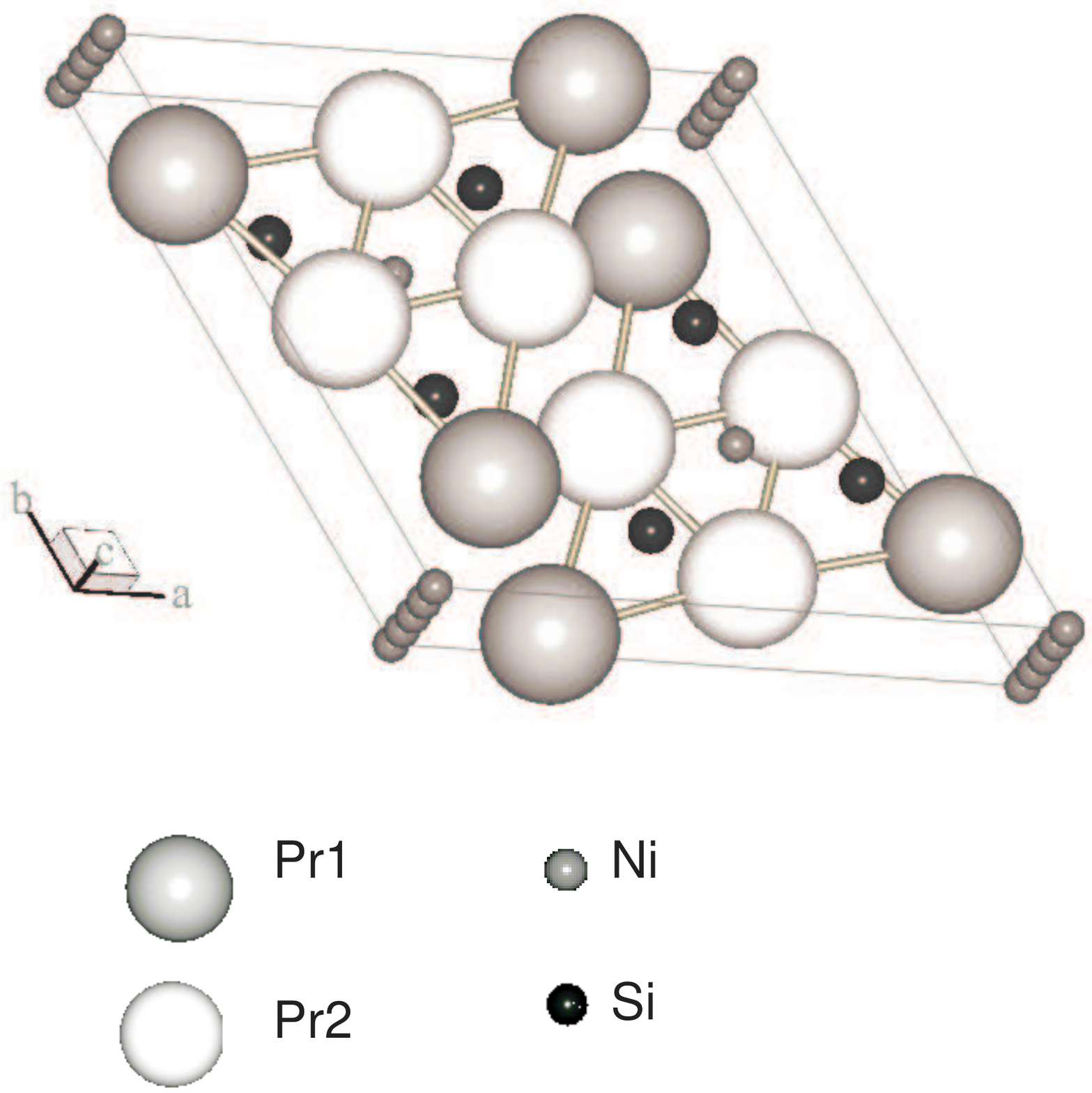}
\begin{center}
\caption{Schematic drawing~\cite{Ozawa04} of the 
Ce$_6$Ni$_2$Si$_3$-type unit cell of \cpd.}\label{Crystalfig}
\end{center}
\end{figure}

In 1984, Parth\'{e} and Chabot\cite{Parthe84} reviewed the crystal structures of ternary 
rare-earth  (R) transition  metal (T) silicide and boride (M) compounds. 
At that time, about 80 different compositions were known. 
A number of the R-T-M compounds can be classified as part of a structure series. 
The majority of these structural series consist of layered structures, and the organization 
of the layers distinguishes members. 
However, there are also at least two R-T-M structure series where the basic building block is a 
triangular prism. These prisms may be assembled into larger prisms in a geometric progression
where each member of this progression represents a unique crystal structure with a specific 
ratio of rare earth to transition metal atoms. The two known series of this type are
described by the formulae $R_{\frac{1}{2}n(n+1)}T_{3(n^2+1)}M_{2n^2+1}$, with as $n=2$ member 
UCo$_5$Si$_3$ 
and   $R_{(n+2)(n+1)}Ni_{n(n-1)+2}Si_{n(n+1)}$, with as $n=2$ member the title compound 
\cpd\ (Fig.~\ref{Crystalfig}).  
In the former series there is only a single rare earth site so the potential for 
competing interactions is small.  In addition, the low concentration of both rare earth 
and transition metal ions is expected to result in weak magnetic interactions and low 
ordering temperatures. On the contrary, the latter series consists of roughly 50 \% 
rare-earth and, moreover,  there are different R local environments, 
see e.\ g.\ Ref.~\onlinecite{Jiles06,Pecharsky03}. 
Furthermore, both the $n=3$ and the $n=4$ 
members are known to order magnetically~\cite{Pecharsky03,Song05a,Song05b, Jiles06}. 

In hexagonal \cpd, with space group P 63/m, the Pr ions occupy two independent 
low-symmetry $6h$ sites, denoted as Pr1 and Pr2 in Fig.~\ref{Crystalfig}. The 
crystal~\cite{Parthe84} structures of the other members of the structure series, 
respectively \cpdf\ and \cpdft, 
have the same space group as \cpd. In \cpdf, Pr ions occupy 3 independent $6h$ sites and
one $2d$ site. In \cpdft, Pr ions occupy 5 independent $6h$ sites. Both \cpdf\ and 
\cpdft\ have Pr ions occupying sites comparable to Pr1 and Pr2 in Fig~\ref{Crystalfig}.
 
Results from polycrystalline samples, for \cpdf\ ($n=3$), or more 
accurately~\cite{Pecharsky03} Pr$_5$Ni$_{1.9}$Si$_3$, and for \cpdft\  ($n=4$) 
indicate that both compounds order ferromagnetically~\cite{Jiles06}, at $\sim$ 50 K, and 
at $\sim$ 60 K, respectively. For both \cpdf\  and \cpdft\ the specific heat shows, besides
the anomaly due to the Curie temperature, another, weaker, anomaly,  at $\sim$27 K and 
at $\sim$33 K, respectively. For both these compounds, at temperatures below the 
low-temperature specific-heat anomaly, the magnetization isotherms show the development 
of a significant coercivity. Moreover, there occurs evidence of metamagnetic-like 
transitions close to 3 T for \cpdf\ at 5 K, and close to 4 T for \cpdft\ at 5 K.

In this paper, we report on the magnetic properties of solution-grown, single-crystalline
\cpd, the $n=2$ member of the aforementioned structure series. 
Results of specific heat, and of extensive anisotropic magnetization measurements are presented. 
Moreover, measurements of the magnetization vector have been used to clarify the 
low-temperature magnetic order, which following crystallographic 
nomenclature~\cite{Toledano01} appears to be 'exotic'.
Finally, the results are discussed within the systematics seen also for the $n=3$ 
and $n=4$ members of the series.

\section{Experimental}
\label{exp}

\begin{figure}[!ht] 
\includegraphics[width=0.45\textwidth,trim=0in 0in 0.0in 0in]{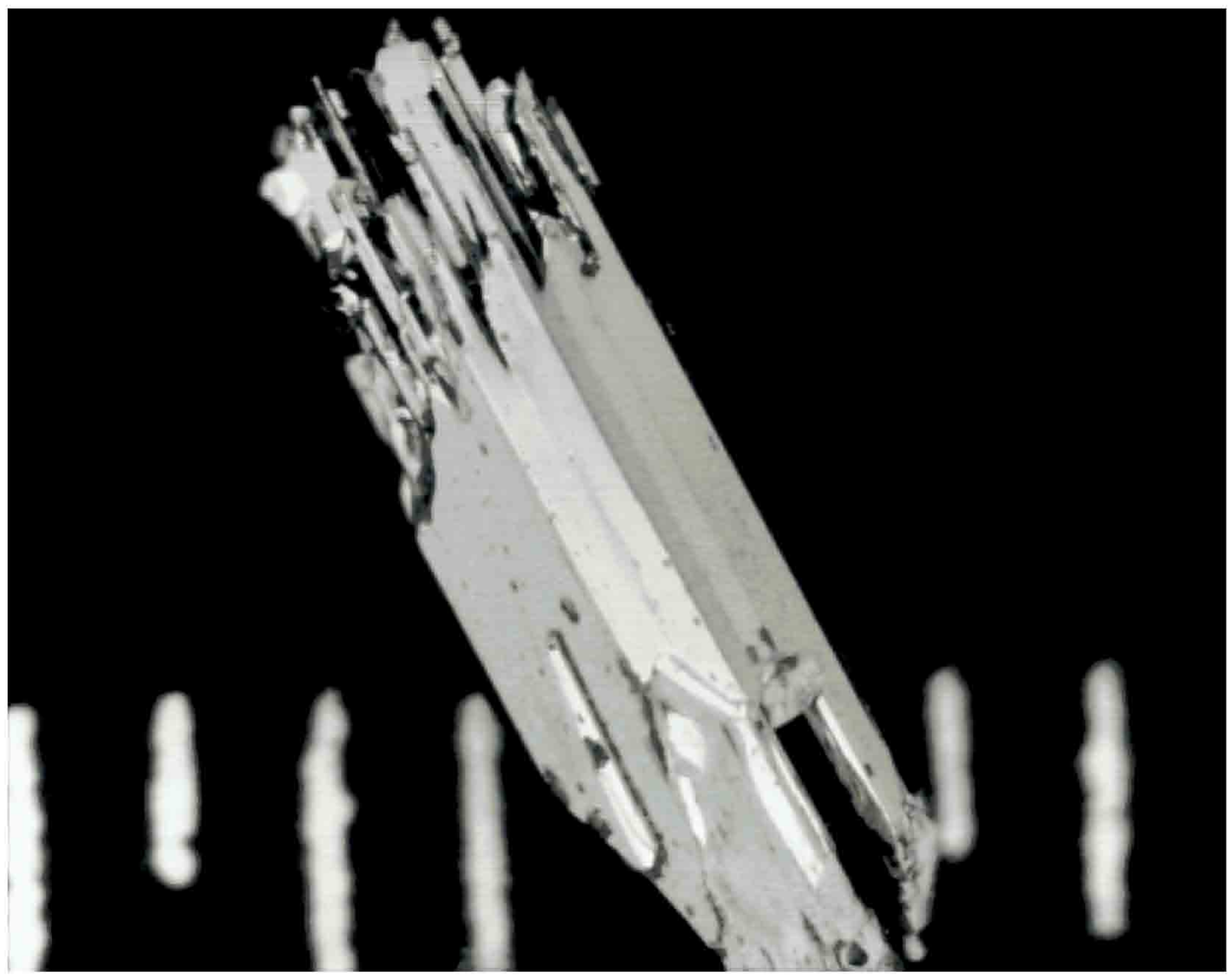}
\begin{center}
\caption{ Photograph of a self-flux grown crystal of \cpd, on a background with a 
\textit{mm}-scale. The crystal axis is the \textit{c} axis, and a [110] facet is 
facing the reader.}\label{photo}
\end{center}
\end{figure}

Single crystals of \cpd\ were grown out of a high-temperature ternary 
solution\cite{Fisk89,Canfield92,Canfield01}. An initial alloy composition and a 
useful temperature range for growth were determined by means of combined DTA and 
growth experiments~\cite{Janssen05}. The initial alloy composition used was 
Pr$_{60}$Ni$_{25}$Si$_{15}$, and the useful temperature range for growth was 
determined to be between $\sim 1000^{\circ}{\rm C}$ and  $\sim 880^{\circ}{\rm C}$. 
The starting elements were sealed in a 3-cap Ta crucible~\cite{Canfield01}, that 
was sealed in an evacuated quartz ampoule. The ampoule was initially heated up to 
$\sim 1200^{\circ}{\rm C}$, to ensure a well-homogenized alloy, cooled to 
$1000^{\circ}{\rm C}$ at $\sim 50^{\circ}{\rm C/h}$, 
and then cooled down to $880^{\circ}{\rm C}$ at $3^{\circ}{\rm C}$/h. The ampoule was taken 
out of the furnace, inverted and centrifuged, resulting in a separation of crystals from 
an excess liquid. Crystals have a hexagonal-prismatic growth habit with faces parallel to 
the [001] crystallographic direction, and normal to the [110] direction~\cite{Yurij}. The 
crystals were up to 10 mm long, and had effective diameters of up to 1 mm. A photograph 
of a \cpd\ crystal is displayed in Fig.~\ref{photo}.

For initial characterization, we measured a powder x-ray diffraction pattern on finely 
ground crystals from the growth yield with a Rigaku Miniflex+ diffractometer employing 
Cu-Ka  radiation. The pattern was analyzed with Rietica~\cite{rietica}, using a Le 
Bail-type~\cite{LeBail88} refinement, and it was indexed according to the space 
group P63/m, with lattice parameters  
a = 11.96(2) \AA\ and c = 4.27(1) \AA. These results are consistent with those for 
isostructural~\cite{Bodak69} Ce$_{6}$Ni$_2$Si$_3$, which has somewhat larger lattice 
parameters ($a$=12.11 \AA, and $c$=4.32 \AA), consistent with lanthanide contraction. 

Specific heat was determined in a Quantum Design physical property measurement system (QD-PPMS)
at temperatures between 2 K and 70 K. Magnetization measurements were performed using 
Quantum Design magnetic property measurement system magnetometers (QD-MPMS), in 
magnetic fields up to 5 T, and at temperatures between 5 K and 300 K. For most of 
the experiments described below, only the magnetization component parallel to the 
applied field was measured for samples aligned with the applied field parallel and perpendicular 
to the hexagonal \ca. To  investigate a possible in-plane magnetic anisotropy the sample 
was aligned with the field perpendicular to the \ca, and rotated by means of a 
horizontal-axis rotator around the \ca.

Generally, a magnetization vector can be decomposed into three perpendicular vector components. 
We can distinguish a longitudinal component, along the applied field direction, here called \ML, 
and transverse components in the plane perpendicular to the applied field, here called \MX\ 
and \MY. 
We used a QD-MPMS-5 system, which was equipped with both a conventional longitudinal 
pick-up-coil system and a transverse pick-up-coil system. Since the magnetometer is 
equipped with one single transverse 
pick-up-coil system, we used the vertical-axis sample rotator to determine both 
the \MX\ and the \MY\ 
components of magnetization. For these vector measurements, the sample was aligned with the 
unique \ca\ at an angle of about 60 degrees with the applied field direction. 

\section{Results}

\begin{figure}[!ht] 
\includegraphics[width=0.45\textwidth,trim=0in 0in 0.0in 0in]{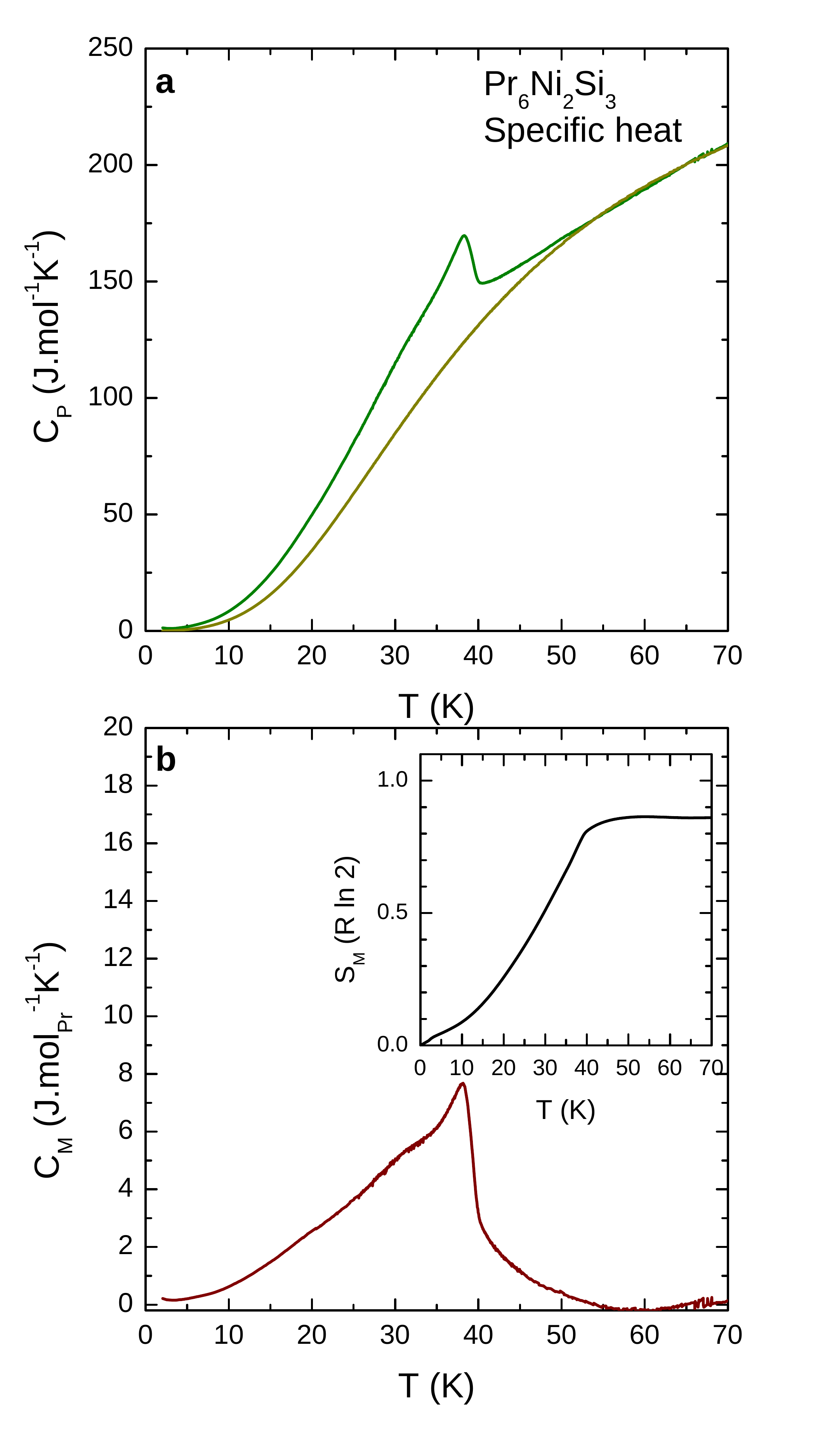}
\begin{center}
\caption{(Color online) \textbf{a} Zero-field specific heat $C_{\rm p}$ as a function of
temperature, also included is an estimate of the lattice specific heat. The mean-field-like 
with an onset near $T$ =
40 K is close to the Curie temperature. \textbf{b} Magnetic specific heat per Pr ion obtained 
from \textbf{a}. Note the shoulder between 30-35 K. The inset shows the estimated magnetic 
entropy to reach values close to $R ln 2$ at the ordering temperature.}\label{CP}
\end{center}
\end{figure}

Temperature-dependent specific heat is presented in Fig.~\ref{CP}\textbf{a}. 
It will be shown below that \cpd\ orders ferromagnetically. Then an onset 
criterion~\cite{Morosan05} for a peak in specific heat can be used. The peak in specific 
heat shows an onset temperature close to 40 K.
A lattice contribution to specific heat, also shown, was estimated according to the 
Debye model, and we obtained a Debye temperature of $\Theta_{\rm D} \approx 165$ K. 
An electronic contribution was ignored. An estimate, Fig.~\ref{CP}\textbf{b}, for 
the magnetic contribution per Pr ion C$_{M}$ was obtained by subtracting this lattice 
contribution from the measured specific heat. Notice that, besides the peak near 40 
K, also a shoulder with a maximum around 30 K can be observed. The temperature-dependent 
magnetic entropy $S_{\rm M}$, estimated by integrating $C_{\rm M}/T$ up to 70 K and 
displayed in the inset of Fig.~\ref{CP}~\textbf{b}, saturates at a value close 
to $R ln 2$ at ~40 K, and then remains approximately constant up to 70 K. This indicates 
that, averaged over the two crystallographically distinct Pr ions, a pair of well-isolated 
singlet states, or a doublet ground state, is responsible for the magnetic order in 
this compound.

\begin{figure}[!ht] 
\includegraphics[width=0.45\textwidth,trim=0in 0in 0.0in 0in]{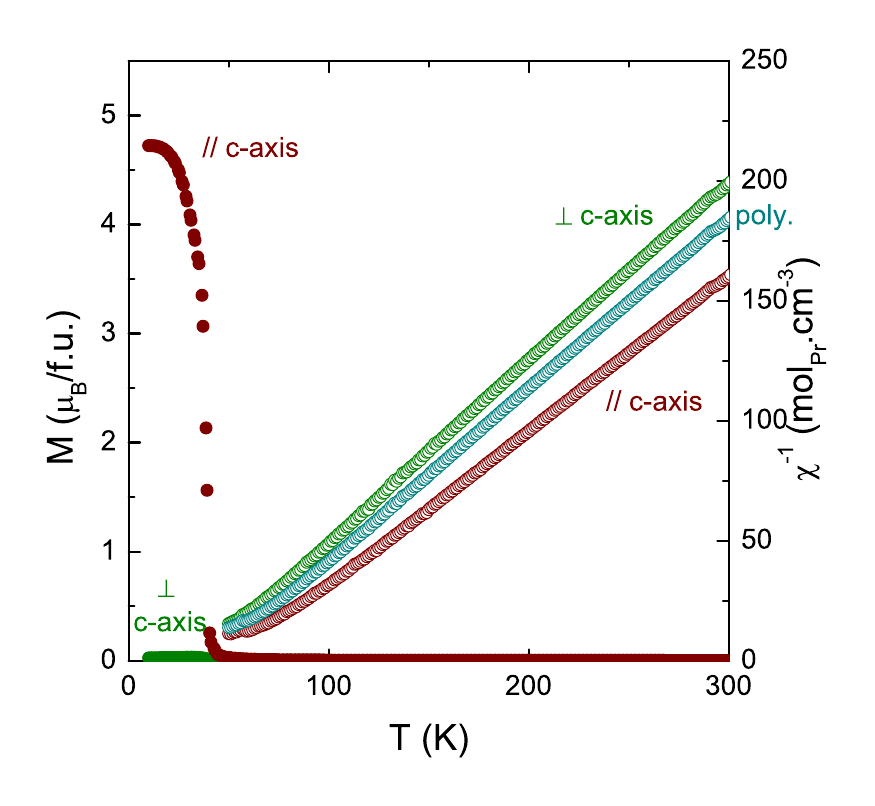}
\begin{center}
\caption{(Color online) Closed circles, left axis: \cpd\ temperature-dependent magnetization 
measured in 0.01 T, both for $H$ // \ca\ (top) 
and for $H$ $\bot$ \ca (bottom). Open circles, right axis: temperature-dependent inverse 
differential susceptibility determined for $H$ // \ca\ (bottom) and $H$ $\bot$ \ca\ 
(top) together with polycrystalline average $\chi^{-1}$ (center).}\label{CW}
\end{center}
\end{figure}

Fig.\ref{CW} shows temperature-dependent magnetization measured upon cooling in a field of 0.01 T 
applied both parallel and perpendicular to the \ca. Two features are immediately 
obvious: the magnetization parallel to the \ca\ indicates a ferromagnetic component // \ca\ 
below $\sim$ 
40 K, and the magnetization $\bot$ \ca\ is much smaller than the magnetization parallel to the 
\ca, especially below 40 K, which indicates that the magnetic anisotropy in this compound is very 
large and favors the moments to align themselves parallel to the \ca. 

Temperature-dependent magnetization for both these sample alignments was determined 
in various fields up to 5 T. From these it was found that above 50 K the magnetization for 
both alignments increases linearly with increasing fields, thus a differential magnetic 
susceptibility $\chi_{\rm diff}=\frac{\Delta M}{\Delta H}$ could be determined. 
A polycrystalline average was obtained by averaging $\chi$ // \ca\ and $\chi$ $\bot$ \ca\ 
according to $\chi_{avg}=(\chi_{//}+2\chi_{\bot})/3$. 

Fig.\ref{CW} also shows the inverted temperature-dependent 
differential magnetic susceptibilities // \ca, $\bot$ \ca\ and 
polycrystalline average. All three are linear, though not parallel, 
with temperature above 100 K, thus can be described by a Curie-Weiss law. 
The effective moments calculated for 
$\chi_{avg}$, $\chi_{//}$, and $\chi_{\bot}$ equal 3.35 $\mu_{\rm B}$/Pr, 
3.50 $\mu_{\rm B}$/Pr, and 3.26 $\mu_{\rm B}$/Pr, which are are all not too far from 
the theoretical free-ion value for Pr (3.58 $\mu_{\rm B}$), thus indicating
that the magnetism in \cpd\ is determined by Pr local magnetic moments. 
However, the fact that the calculated effective moments are different for the different 
crystallographic directions is an indication of a substantial crystal-field splitting 
of the 2J+1 levels of the Pr-4f shell, which is still noticable at room temperature.  
Weiss temperatures $\Theta$ of 41 K, 53 K, and 33 K were found for $\Theta_{avg}$, 
$\Theta_{//}$, and $\Theta_{\bot}$, respectively, positive values which are reasonably 
close to $T_{\mathrm C}$, and which indicate (mainly) ferromagnetic interactions between the Pr
moments.

\subsection{$H$ // \ca}

\begin{figure}[!ht] 
\includegraphics[width=0.45\textwidth,trim=0in 0in 0.0in 0in]{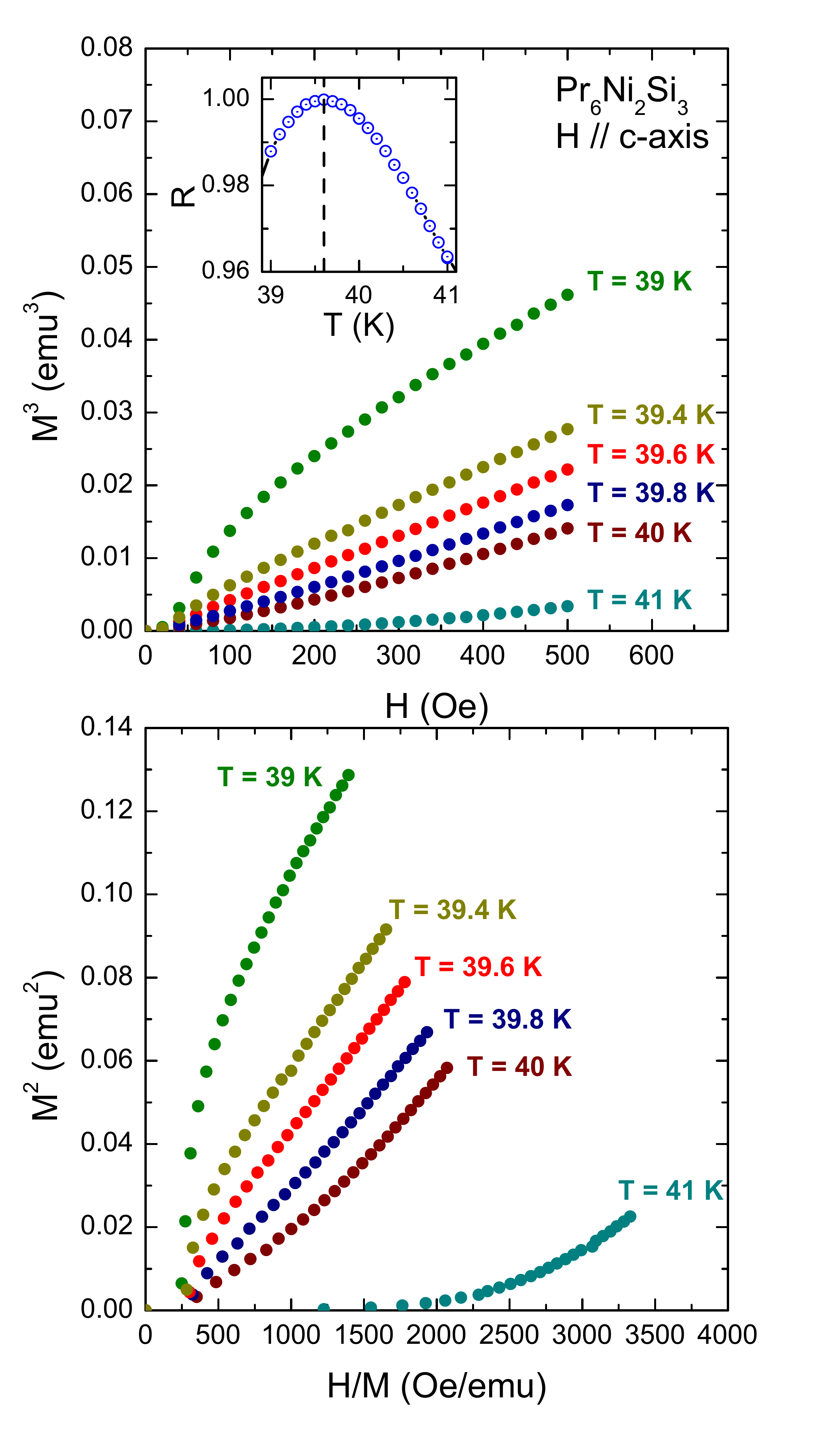}
\begin{center}
\caption{(Color online) \cpd\ Arrott plots for $H$ // \ca. The upper panel shows field ($H$) 
dependent magnetization ($M^3$) for representative temperatures. 
The inset shows the $T$-dependent regression factor $R$ for 
a linear fit through $M^3$($T$), indicating the Curie temperature $T_{\rm C} = 39.6 (1)$ K. 
The lower panel shows conventional Arrott plots, of $M^2$ vs. $H/M$ of the same data sets.}\label{arrott}
\end{center}
\end{figure}

The Curie temperature for the onset of ferromagnetic order // \ca\ can be determined from Arrott 
plots~\cite{Arrott57}. Fig.~\ref{arrott} shows such plots for \cpd, obtained from magnetization 
isotherms $H$ // \ca\ in fields up to 500 Oe. In Arrott's original paper~\cite{Arrott57}, plots 
of $M^3$ vs $H$ are used rather than the more conventional $M^2$ vs $H/M$. 
According to Arrott's criterion, precisely at $T_{\rm C}$, the magnetic susceptibility $\chi$ 
tends to infinity, causing terms of $M^3$ to be dominant at low enough $H$. In other words, 
the Curie temperature is at that  temperature where $H$-dependent $M^3$ is exactly linear 
starting at $H$ = 0. A criterion for the linearity of a fit line is given by the 
regression factor $R$, which ranges between 0 and 1, where 1 indicates a perfect line. 
The top panel of Fig.~\ref{arrott} shows $M^3$ vs $H$ taken at temperatures between 39 K 
and 40 K, 
and the inset to that panel shows the temperature-dependent R of such linear fits. From this the 
Curie temperature is determined as $T_{\rm C}$ =39.6 (1) K. This value agrees very well with 
the above 
obtained value for the onset of the specific heat peak. For completeness, a more conventional 
Arrott plot of $M^2$ vs $H/M$ is displayed in the bottom panel of Fig.~\ref{arrott}. As expected, 
the curve taken at 39.6 K extrapolated to zero $M^2$ is closest to intercepting the $H/M$ axis. 

\begin{figure}[!ht] 
\includegraphics[width=0.45\textwidth,trim=0in 0in 0.0in 0in]{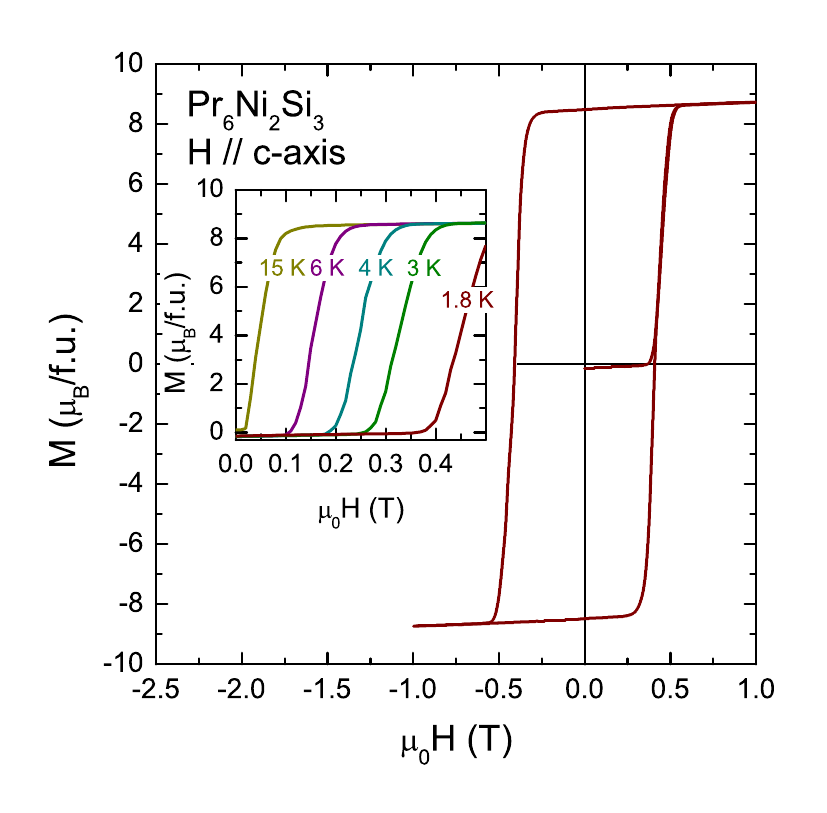}
\begin{center}
\caption{(Color online) \cpd\ zero-field-cooled magnetization loop for $H$ // \ca\ at 1.8 K. 
The inset shows the development of virgin magnetization with temperature, demonstrating the 
coercive behavior.}\label{MHparc1p8}
\end{center}
\end{figure}

Hysteretic behavior of \cpd\ at 1.8 K for $H$ // \ca\ is shown in Fig.~\ref{MHparc1p8}. 
These results were obtained by cooling the aligned crystal in zero field and measuring 
first the virgin magnetization curve. The observed presence of high coercivity in 
conjunction with the small slope of the virgin curve can be taken as a signature of 
the presence of narrow domain walls~\cite{Barbara71, Egami71, Broek71, Barbara76, Tegus01}. 
Such narrow walls can be strongly pinned by magnetic obstacles of atomic dimensions. 
The strong increase of the magnetization on the virgin curve at $H_{\rm P}$ = 0.45 T marks the 
propagation field $H_{\rm P}$ at which the external field is able to detach the narrow walls 
from the pinning sites. At higher fields the walls are removed from the crystal. 
Upon decreasing the fields from 1 T into the region of negative fields, 
reversed domains and domain walls can nucleate but the movement of these walls is 
impeded by the pinning sites so that the reversed domains cannot grow. This becomes 
possible again only for negative fields equal in magnitude to $H_{\rm P}$,
causing the absolute value of the coercive field to be equal to the 
propagation field, $H_{\rm C} = H_{\rm P}$.  

Because of the strong hysteresis, the spontaneous magnetization $M_{\rm S}$ at 1.8 K can 
directly be obtained from this Figure. 
It equals about 8.5 $\mu_{\rm B}$/f.u., amounting to about 1.4 $\mu_{\rm B}$/Pr ion, which 
is much lower than the free-ion value of 3.2 $\mu_{\rm B}$/Pr, which may be related to strong 
crystal-electric field effects and the low point symmetry of both Pr-crystallographic sites. 

The temperature dependence of the coercive field is demonstrated in the inset of 
Fig.~\ref{MHparc1p8}, which shows virgin $M$($H$) 
at different temperatures below 15 K. Note that besides having a different coercive 
field these curves overlap, so $M_{\rm S}$ 
is almost temperature independent below 15 K. 

\begin{figure}[!ht] 
\includegraphics[width=0.45\textwidth,trim=0in 0in 0.0in 0in]{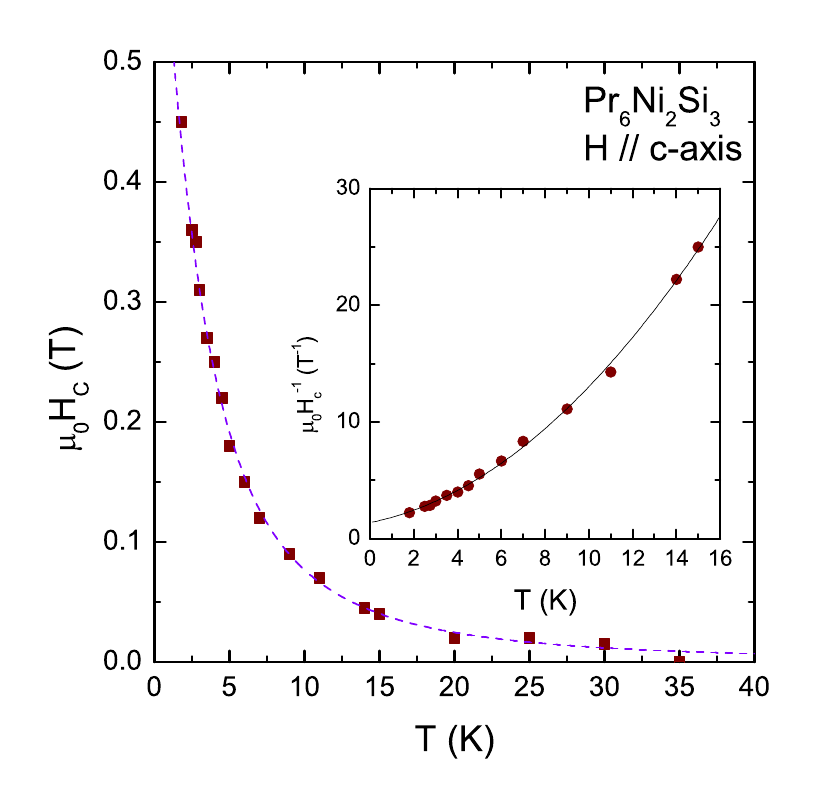}
\begin{center}
\caption{(Color online) \cpd\ temperature-dependent coercive field for $H$ // \ca. The line 
is obtained as descibed in the text. }\label{HCT}
\end{center}
\end{figure}

Measurements at various temperatures below $T_{\rm C}$ indicate the same behavior as at 1.8 K,
but with a strongly temperature-dependent $H_{\rm C}$ Fig.~\ref{HCT} shows 
temperature-dependent $H_{\rm C}$. According to a model proposed by Barbara and 
Uehara~\cite{Barbara76}, the temperature dependence of coercivity can be described as:

\begin{equation}
H_{\rm C}^{-1} (T) = H_{\rm C}(0)^{-1} + \alpha T,\label{hcteq}
\end{equation}

where $\alpha$ is proportional to the spontaneous magnetization divided by the domain-wall 
energy, $\alpha \propto M_{\rm S}/\gamma^2$. $\gamma^2$ in turn is proportional to the product
 of the average exchange energy and the average anisotropy energy.  The inset of 
Fig.~\ref{HCT} shows temperature-dependent $H\_{\rm C}^{-1}$ at temperatures between 1.8 K 
and 15 K. A very good fit to this line is given by a second-order polynomial, and when 
comparing to Eq.~\ref{hcteq}, this means that $\alpha$ is linearly dependent on $T$. Values 
found are: $H_{\rm C} (0)^{-1}$ = 1.4(3) T$^{-1}$, and $\alpha (T) = 0.37(9) + 0.079(6) T$, 
leading to a zero-temperature $H_{\rm C}$ of 0.71 T. The dotted line in the main 
Fig.~\ref{HCT} was calculated using these values.

Since $M\_{\rm S}$ is constant below 15 K, $M_{\rm S}$ is not expecting to contribute to 
variations of $\alpha (T)$ below 15 K. Therefore, the variations in $\alpha$ below 15 K 
have to be proportional to the variations of $1/\gamma^2$, the inverted product of the 
average exchange energy and the average anisotropy energy. Both the exchange energy 
and the anisotropy energy may be expected to decrease with increasing temperature, leading 
to $\alpha$ growing with increasing temperature.
 
\subsection{$H$ $\bot$ \ca}

\begin{figure}[!ht] 
\includegraphics[width=0.45\textwidth,trim=0in 0in 0.0in 0in]{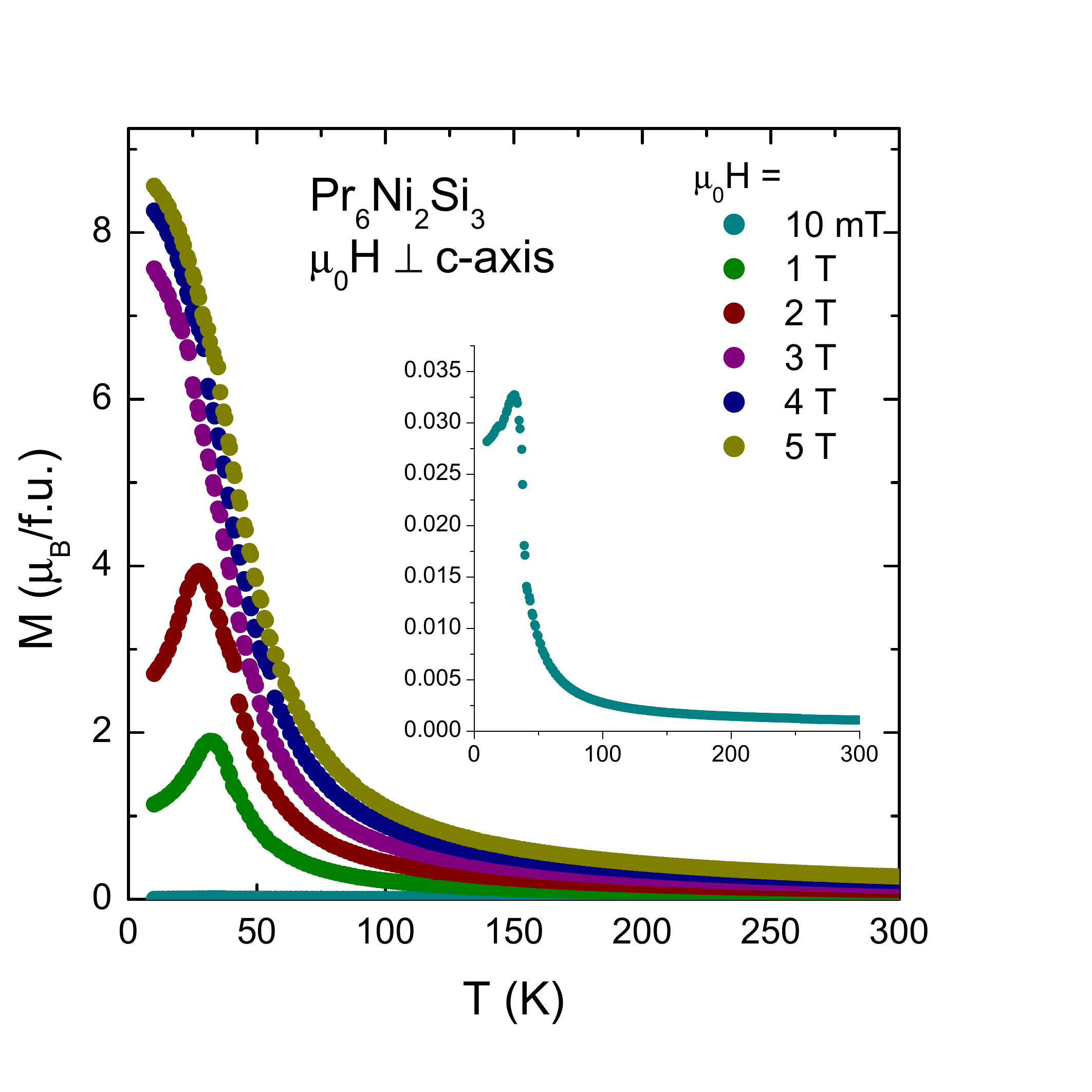}
\begin{center}
\caption{(Color online) Temperature-dependent magnetization for $H$ $\bot$ \ca\ measured in 
various fields up to 5 T. Note the maxima for 0.01 ($\approx$ 32 K), 1 ($\approx$ 32 K, and 
2 T ($\approx$ 27 K.}\label{mtperp}
\end{center}
\end{figure}

Fig.~\ref{mtperp} shows temperature-dependent magnetization for $H \bot$ \ca\ measured, 
with temperature decreasing,
in various fields up to 5 T. In 0.01, 1 and 2 T, the magnetization shows a maximum, 
which appears close to 32 K, for 0.01 T and 1 T, and close to 27 K for 2 T. In 3 T 
and higher, no maximum is observed. Such behavior, a maximum 
in temperature-dependent magnetization, which shifts to lower temperature with increasing 
field strengths, is common in antiferromagnets~\cite{DeJongh74}. Note also that the 
maximum in the 0.01 T curve occurs at a temperature close to the 
32 K shoulder in specific heat. 

\begin{figure}[!ht] 
\includegraphics[width=0.45\textwidth,trim=0in 0in 0.0in 0in]{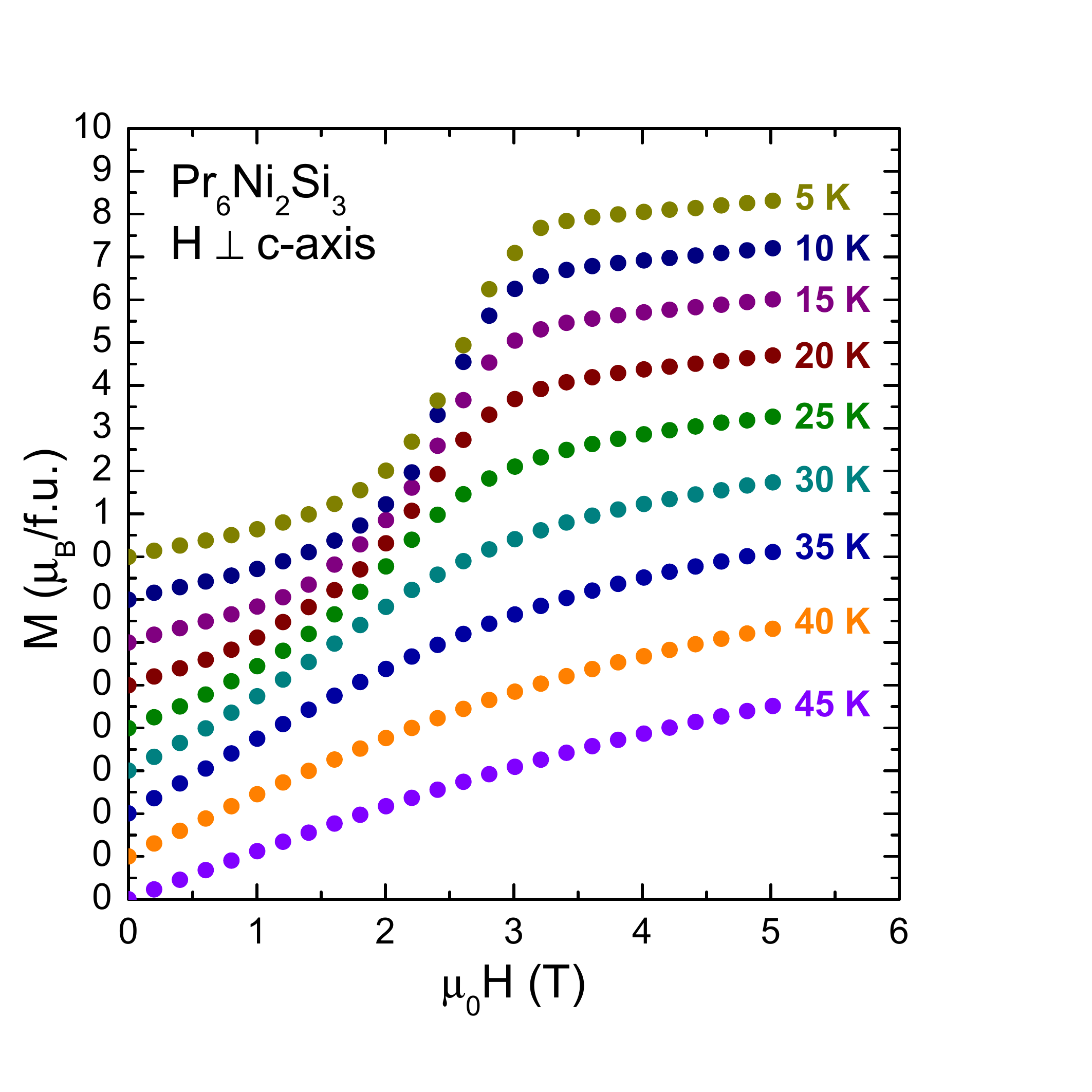}
\begin{center}
\caption{(Color online) Field-dependent magnetization of \cpd\ at various temperatures measured 
with $H$ $\bot$ \ca. For clarity, the curves have been shifted by 1 $\mu_{B}$/f.u. with respect 
to one another.}\label{mhperp}
\end{center}
\end{figure}

Field-dependent magnetization isotherms for $H \bot$ \ca\ at temperatures between 5 K and 45 K 
are shown in Fig.~\ref{mhperp}. At 5 K, starting in zero field, the magnetization starts at 
zero and first increases weakly and linearly with increasing field. Close to 2 T, the 
magnetization starts to increase much faster with increasing field, a process which 
ends close to 3 T, above which the magnetization continues to increase linearly with 
increasing field, at a slope similar to the slope in low field. As temperature increases, 
this process becomes less pronounced, resulting in a weakly s-shaped magnetization at 30 K, 
and a featureless magnetization at 35 K and higher. Also these results are consistent with 
the magnetization of a simple antiferromagnet, and the magnetization process could be 
interpreted as due to a spin flop~\cite{DeJongh74}.

\begin{figure}[!ht] 
\includegraphics[width=0.45\textwidth,trim=0in 0in 0.0in 0in]{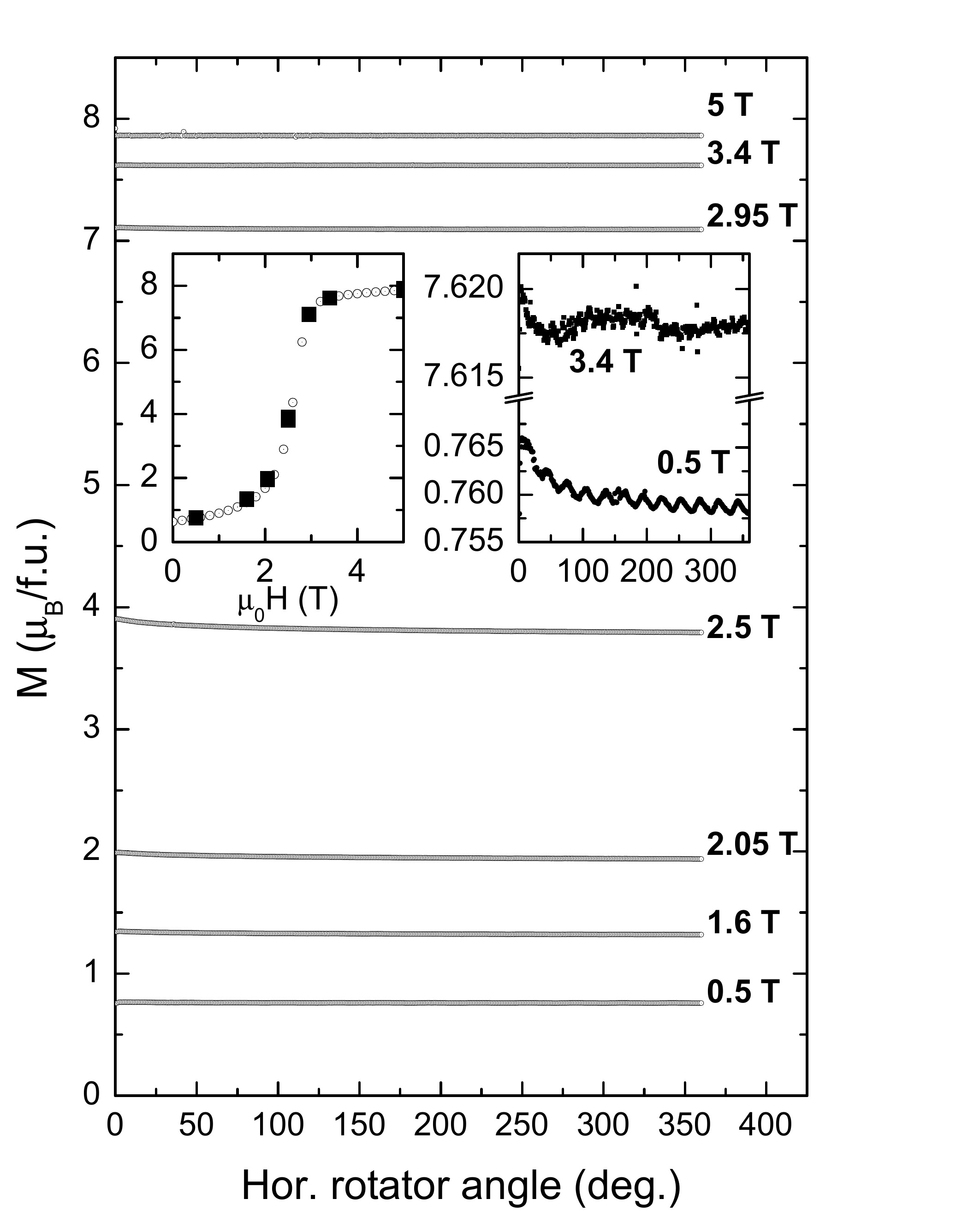}
\begin{center}
\caption{\cpd\ horizontal-angle dependent magnetization, measured at 5 K, in various fields 
(H $\bot$ \ca) up to 5 T. The left inset shows all data as a function of $H$, superimposed 
on a denser dataset measured at zero angle. The right inset shows a zoom in of the 0.5 T 
data (bottom) and the 3.4 T data (top).}\label{mhrot}
\end{center}
\end{figure}

To determine magnetic anisotropy in the plane perpendicular to the \ca\ we more precisely 
measured the magnetization at 5 K for the field $\bot$ \ca\ with a horizontal rotator, 
rotating the sample around the \ca. In a hexagonal system with a very strong in-plane 
anisotropy, the in-plane magnetization may vary by as much as $1 - cos 30^\circ \sim 15 \%$.   
Fig.~\ref{mhrot} shows the magnetization in various fields up to 5 T, measured at horizontal 
rotator angles between 0 and 360$^\circ$. For all chosen field strengths, the variation in 
magnetization is smaller that the symbols used for the figure. The in-plane anisotropy 
of \cpd\ in fields up to 5 T is thus very small, which is exemplified in the left inset,
by plotting all thus measured magnetization values on a more densely measured field-dependent 
magnetization curve, measured at zero angle. Strongly zoomed in, see right inset of 
Fig.~\ref{mhrot}, e\. g\. the magnetization at 0.5 T shows a very weak and 12 fold 
variation, with an amplitude of variation of about 0.1 \%. This variation disappears 
in higher fields: the zoomed in angular dependent magnetization measured in 3.4 T 
shows no such variation. 
This 12 fold variation may be related to the two crystallographically distinct magnetic 
Pr sites in the unit cell of \cpd. 

\subsection{$H$ applied at 60$^\circ$ from the \ca}

\begin{figure}[!ht] 
\includegraphics[width=0.45\textwidth,trim=0in 0in 0.0in 0in]{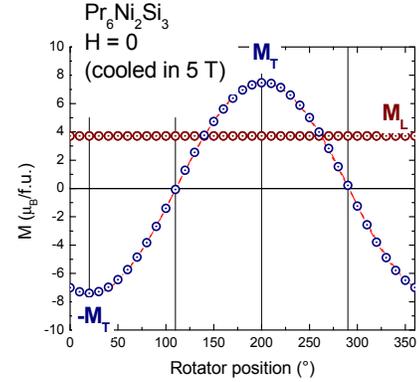}
\begin{center}
\caption{\cpd\ zero-field magnetization, measured at 5 K both parallel to the applied field 
$H$ ($M_{\rm L}$) and perpendicular to it ($M_{\rm T}$) as a function of (vertical) rotator 
position. This was measured on a field-cooled sample which was mounted with its \ca\ at 
an angle of approx. 60 degrees with $H$.}\label{MZFtilted}
\end{center}
\end{figure}

The crystal was mounted and centered in a straw such that the applied field $H$ made an angle 
of $\approx 60^\circ$ with the \ca. The very weak in-plane anisotropy, Fig.~\ref{mhrot}, 
made it clear that no particular attention to the planar direction closest to the applied 
field was necessary, which is not true if the in-plane anisotropy is strong\cite{Janssen02}. 
The sample was then cooled in an applied field of 5 T from room temperature to 5 K, at 
which temperature the field was removed. Vertical-rotator angle-dependent zero-field 
magnetization measured both parallel (longitudinal magnetization $M_{\rm L}$) and 
perpendicular to $H$ (transverse magnetization $M_{\rm T}$) are shown in 
Fig.~\ref{MZFtilted}. Whereas the measured $M_{\rm L}$ is vertical-angle independent, 
$M_{\rm T}$ is determined as the amplitude of the measured transverse magnetization 
cosine. 
We thus find for $M_{\rm T}$ a value of 
$\approx 7.5 \mu_{\rm B}$/f.u. and for $M_{\rm L}$ 3.8 $\mu_{\rm B}$/f.u. The angle of the 
magnetic moment with the applied field is given by
$tan^{-1} \frac{M_{\rm T}}{M_{\rm L}}=63^\circ$, which is in good agreement with the angle 
at which the crystal axis was mounted in the sample holder. The size of the moment
vector $|M| = \sqrt{M_{\rm L}^2+M_{\rm T}^2} = 8.4 \mu_{\rm B}$/f.u. is in excellent 
agreement with the magnetization moment found at 5 K for $H$ // \ca.

\begin{figure}[!ht] 
\includegraphics[width=0.45\textwidth,trim=0in 0in 0.0in 0in]{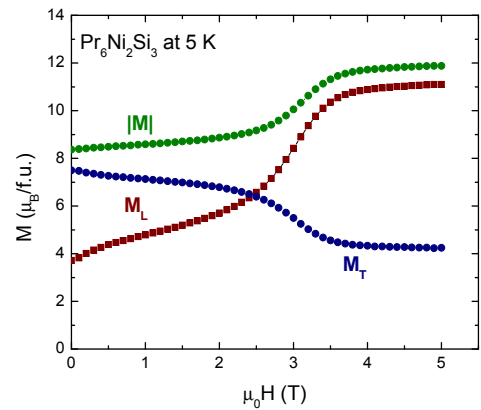}
\begin{center}
\caption{\cpd\ field-dependent magnetization, measured at 5 K both parallel (\ML) and 
perpendicular (\MT) to the applied field, resulting in a vector-summed magnetization 
$|M|$.}\label{MHtot5K}
\end{center}
\end{figure}

Field-dependent magnetization was determined by measuring both $M_{\rm L}$ and $M_{\rm T}$ at 
5 K in various $H$ up to 5 T. $M_{\rm L}$ was determined in the conventional way. Full 
rotations of the vertical rotator, similar to the measurement shown in Fig.~\ref{MZFtilted}, 
were made to determine $M_{\rm T}$. In this way, we verified that the magnetic moment does 
not rotate in the plane perpendicular to the magnetic field, which was a distinct 
possibility~\cite{Janssen02}. The results are shown in Fig~\ref{MHtot5K}. As expected, 
\ML\ increases uniformly with increasing $H$, and the spin-flop-like transition starts 
close to 2.5 T, a field some 15\% (=1/sin 60$^\circ$) higher than for the measurement 
shown in Fig.~\ref{mhperp}. At the same time, \MT\ decreases uniformly with increasing 
fields, which generally indicates a rotation of the magnetic moment towards H. The 
spin-flop-like transition for this magnetization-vector component mimics the one for \ML, 
but as a stronger decrease rather than an increase. The amount by which the magnetization
vector decreases for \MT\ due to the spin-flop-like transition may seem small compared to 
the increase observed for \ML. This is clarified examining the total magnetization, the vector 
sum $|M|$, also shown in Fig.~\ref{MHtot5K}. $|M|$ shows a small, linear increase with increasing 
field strenghts up to $\sim$ 2.5 T, and increases even further during the spin-flop-like 
transition, up to about 3.5 T, above which it further increases slightly and linearly. 
This anomalous change in length of the magnetization vector is an immediate indication 
that the metamagnetic spin-flop-like transition for $H$ $\bot$ \ca\ is not due to a 
simple rotation of the magnetization vector~\cite{Janssen02}.

\begin{figure}[!ht] 
\includegraphics[width=0.45\textwidth,trim=1in 5in 4in 1in]{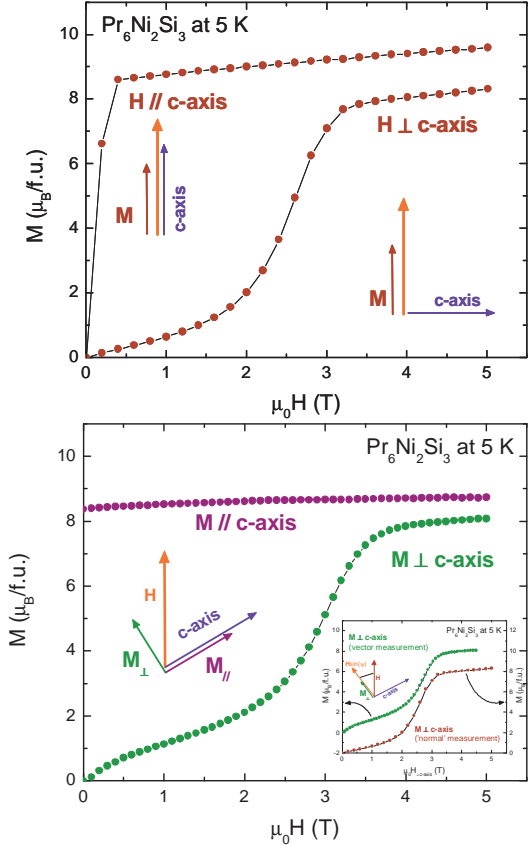}
\begin{center}
\caption{\cpd\ field-dependent magnetization, measured at 5 K both parallel (\ML) and 
perpendicular (\MT) to the applied field, resulting in a vector-summed magnetization 
$|M|$.}\label{MHprojected}
\end{center}
\end{figure}

The lower panel of Fig.~\ref{MHprojected} shows the measured field-dependent magnetization 
vector with the field applied at $\sim 60^\circ$ from \ca\ (Fig.~\ref{MHtot5K}) 
decomposed~\cite{Janssen02} in a component // \ca\ and a component $\bot$ \ca. 
For comparison, field-dependent longitudinal magnetization measured with $H$ applied  // \ca\
and with $H$ applied $\bot$ \ca\ are shown in the upper panel of Fig.~\ref{MHprojected}. 
In both the lower and uppper panel, we show an almost constant magnetization // \ca, 
whereas the magnetization $\bot$ shows a spin-flop like transition. The inset of the 
lower panel of Fig.~\ref{MHprojected} shows the magnetization $\bot$ \ca\ obtained by 
vector magnetometry with $H$ projected $\bot$ \ca, compared to the longitudinal 
magnetization from the upper panel of Fig.~\ref{MHprojected}. Thus the spin-flop like 
process $\bot$ \ca\ occurs independent from the magnetization // \ca, and is only due 
to the $H$ $\bot$ \ca.  

\section{Discussion and conclusions}

The experimental results shown above indicate quite clearly that the magnetic order in the 
intermetallic compound \cpd\ has both ferromagnetic and antiferromagnetic components, 
which may be callled 'exotic' following proposed nomenclature~\cite{Toledano01}. 
As described below, it may well be that the ferromagnetic order mainly involves 
one of the two Pr sites, and the antiferromagnetic order the other.
In our view, two magnetic phase transitions may be discerned, possibly with 
different propagation vectors, q=0 for the ferromagnetic order and another one for 
the antiferomagnetic order.
The first transition occurs at $T_{\rm C}$ = 39.6 K, 
where Pr moments order ferromagnetically // \ca, with a spontaneous magnetization 
substantially lower than the theoretical free-ion moment for Pr. This transition is 
evidenced by a 
clear specific heat anomaly and by Arrott plots of $M (H)$ for $H$ // \ca. The second 
magnetic transition is due to antiferromagnetic order $\bot$ \ca\ and shows as a 
weak shoulder in specific heat close to 32 K, and in low fields $\bot$ \ca\ by $M (T)$, 
where a peak occurs close to 32 K. Furthermore, spin-flop-like transitions are 
only clearly observed below 32 K.

The ferromagnetic order // \ca\ is further corroborated by the strong and strongly 
temperature-dependent coercivity that occurs for $M$ // \ca, which shows itself in 
nearly square magnetization loops at low temperatures. Such loops occur in ferromagnets 
with narrow domain walls which are strongly pinned, on obstacles of atomic size. The 
coercivity in \cpd\ becomes stronger with decreasing temperatures, which may be expected 
for any coercive magnet. We find no clear evidence that the behavior of the coercive 
field is related to the development of antiferromagnetic order $\bot$ \ca. The cause of 
the narrow domain walls, which are not usually observed in pure single crystals is not 
presently clear, but may be related to the instrinsic crystallographic disorder we have 
found~\cite{Yurij}.

The antiferromagnetic order $\bot$ \ca\ in turn is corroborated by a magnetization process, 
similar to a spin-flop transition, which starts close to 2 T at 2 K for $H \bot$ \ca. That 
this magnetization process is only due to magnetization for $H \bot$ \ca\ is evidenced by 
measurements of the magnetization vector for $H$ applied at an angle of 60$^\circ$ with 
the \ca. Although the fact that $M_{\rm S}$ // \ca\ is substantially smaller than the 
full free-ion Pr moment enables additional magnetic order, there is no evidence that 
here these two magnetic orderings // \ca\ and $\bot$ \ca\ are linked.  
 
The magnetic properties of single-crystalline \cpd\ is consistent with the magnetic 
properties of the other members of the structure series, \cpdf\ and \cpdft, which order 
also ferromagnetially at 50 and at 60 K, respectively. Both these compounds also show,
besides the Curie-temperature anomaly, shoulders in specific heat, at 27 K and at 33 K, 
respectively, below which temperatures for both these compounds in the, polycrystalline, 
magnetization metamagnetic-like behavior appears. 

Preliminary studies on single crystals~\cite{Janssen06} of these compounds indicate that 
they also order ferromagnetically // \ca\ and antiferromagnetically $\bot$ \ca. 
Furthermore, a preliminary neutron powder diffraction~\cite{Llobet05} measurement 
of \cpdf, indicated that Pr moments 
order ferromagnetically // \ca, and the moment on the site comparable to Pr1 in 
Fig.~\ref{Crystalfig} has a small moment // \ca\ compared to the other Pr sites. 
Also, a separate incommensurate diffraction peak was found at low temperature, 
which we assume is due to antiferromagnetic order mainly on this crystallographic site. 

Concluding, the above presented results on \cpd\ indicate that its ordered state manifests
both ferromagnetic and antiferromagnetic components. The results are not only consistent
with results obtained on other members of the structure series of which it forms part, 
but also clarifies these.  

\label{conc}

\section{Acknowledgments} \label{ack} We are indebted to S. L. Bud'ko, Y. Mozharivskyj, and 
J. Frederick for valuable discussions and for help with the experiments.
Work at the Ames Laboratory was supported by the Department of Energy, Basic Energy Sciences under Contract No. DE-AC02-07CH11358.

\end{document}